\documentclass[twocolumn,prl]{revtex4}
\usepackage{jpc}
\usepackage{graphics}
\usepackage{amssymb}
\usepackage{amsmath}

\def\cU{{\cal{U}}}

\begin{document}


\title{ Modelling a suspended nanotube oscillator}

\author{H. \"Ust\"unel$^*$, D. Roundy and T. A. Arias\\
\emph{Department of Physics, Cornell University, Ithaca, NY 14853}\\
\small{$^*$corresponding author: \tt{hande@physics.cornell.edu}}}

\begin{abstract}
We present a general study of oscillations in suspended
one-dimensional elastic systems clamped at each end, exploring a wide
range of slack (excess length) and downward external forces. Our
results apply directly to recent experiments in nanotube and silicon
nanowire oscillators.  We find the behavior to simplify in three
well-defined regimes which we present in a dimensionless phase
diagram.  The frequencies of vibration of such systems are found to be
extremely sensitive to slack.
\end{abstract}


\maketitle

Vibrations of {\em one-dimensional} systems (i.e. systems with
cross-sectional dimension much smaller than their length) suspended under
the influence of a downward force have long been of interest in the context
of such applications as beams, cables supporting suspension bridges and
ship moorings~\cite{Irvine,AIAA,mooring}.  Such systems display a wide
range of behavior, depending on the amount of slack present in the system,
the downward force, and the aspect ratio.  Previous studies, which have
largely been analytical, have been restricted to certain limiting cases of
these parameters.  Recently work on oscillating nanoscale systems, such as
carbon nanotubes\cite{Delft} and silicon
nanowires\cite{Carr}, has opened the
possibility of experimentally exploring such vibrations in entirely new
regimes.  In the present work, we study numerically the oscillations of a
one dimensional elastic system over an extensive range of both the slack
and force parameters, providing insight into the physics which separates
this parameter space into three distinct regimes of behavior.

\begin{figure}
\begin{center}
\begin{tabular}{c}
\scalebox{0.30}{\includegraphics{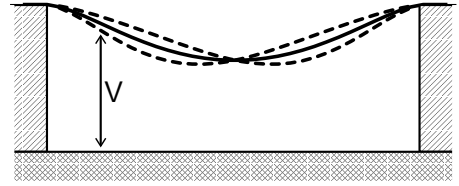}} \\
(a) \\
\scalebox{0.30}{\includegraphics{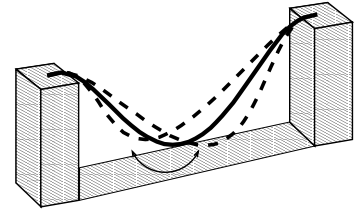}} \\
(b) \\
\end{tabular}
\end{center}
\caption{(a) Suspended nanotube under the action of a gate voltage
  $V$. (b) \emph {Jump-rope} mode}

\label{fig:experiment}
\end{figure}

The treatment below is entirely general with illustrative examples taken
from parameters relevant to carbon nanotubes.  Since their discovery in
1991~\cite{Iijima}, nanotubes have found many applications in device
technology due to their small size, robust structure and superior elastic
properties~\cite{Stevens, Keren, Shimada, Yang}. Many of these applications
involve the use of nanotubes as mechanical oscillators, making theoretical
understanding of the vibrational properties of nanotubes in various
geometries of current interest\cite{Delft,cantilever}.  Recent
experiments~\cite{Nature} have studied the behavior of the transverse
vibrations of a suspended nanotube clamped at both ends, under the action
of a downward force, as sketched in Figure~\ref{fig:experiment}.  These
suspended nanotubes generally have around 1\% \emph{slack}, denoted in this
work by $s$, which we define to be the ratio of the excess length of the
tube to the distance between clamping points.

{\em Analytic model and computational techniques ---} The potential
energy of a one-dimensional elastic continuum under a uniform downward
force is
\begin{equation}
\cU=\frac{1}{2}\int_0^L \left[Eu^2(l) + \frac{F}{R^2(l)} + fz(l)\right] dl
\label{energy}
\end{equation}
Here, $l$ represents distance along the system of unstretched length $L$;
$u(l)$, $R(l)$ and $z(l)$ represent the local strain, radius of curvature
and vertical displacement, respectively; $E$ and $F$ are the extensional
and flexural rigidities; and $f$ is the downward force per unit length.
For single-walled nanotubes with diameters $d > 1$~nm, the
rigidities we obtain using the Tersoff-Brenner empirical
potential~\cite{Brenner} are, to within 0.5\%,

\begin{align}
E&=1.09 \cdot d  & {\text {(TPa $\cdot$ nm$^2$)}}  \notag \\
F&=Ed^2/8=0.14\cdot d^3  &{\text {(TPa $\cdot$ nm$^4$)}},   
\label{EandF}
\end{align}
regardless of the chirality of the tube~\cite{youngs}.  We note that
essentially the same values of $E$ and $F$ can be obtained using the
in-plane elastic modulus of graphite.

To evaluate the potential energy (Equation~\ref{energy}), we begin with
$\mathbf{r}(l)$, the {\em vector} location of the segment at position $l$
from the end of the tube.  The local strain is, then,
\begin{equation}
u(l) \equiv \left|\mathbf{r}'\right| - 1 
\end{equation}
and inverse radius of curvature is
\begin{equation}
\frac{1}{R} \equiv \frac{\left| (1-\hat{t}\hat{t}\cdot)\mathbf{r}''\right| }{ \left|
  \mathbf{r}' \right|^2 },
\end{equation}
where prime indicates differentiation with respect to $l$, and the
tangent vector $\hat{t}\equiv \mathbf{r}' / \left| \mathbf{r}'
\right|$.

For a given force and slack, we discretize $\mathbf{r}(l)$ along $l$ with a
number of points (typically 30) sufficient to render all plots shown below
visually indistinguishable under increasing resolution.  To find the
equilibrium configuration of the system under the action of the external
force, we then relax the system using a conjugate gradients algorithm with
forces computed using finite differences.  Finally, we compute the force
constant matrix $K_{ij}\equiv\frac{\partial ^2 {\cal {U}}}{\partial x_i
\partial x_j}$ and diagonalize it to obtain frequencies $\nu$ while
assigning a mass $\mu \, \Delta l$ to each discretized point, where $\mu$
and $\Delta l$ are the mass per unit length of the system and the distance
between sample points respectively.

For a one-dimensional system, slack $s$, length $L$ and
downward force per unit length $f$ can be regarded as \emph{external
parameters} that can be changed independently. Mass density $\mu$ and
rigidities $E$ and $F$ on the other hand, all depend on the cross-sectional
geometry of the system and can therefore be considered \emph{internal
parameters}. To make our discussion applicable to a general one-dimensional
system, we present our results in units scaled by appropriate combinations
of these basic parameters.  In addition, we give results in physical units
for the concrete system of a nanotube of diameter 2~nm with length
$L=1.75$~$\mu$m.  For frequency plots, besides the physical frequency
$\nu$, we plot a dimensionless frequency $\tilde \nu \equiv \nu \sqrt{(\mu
L^4/F)}$.

\begin{figure}
\begin{center}
  \begin{tabular}{c }
    \scalebox{0.30}{\includegraphics{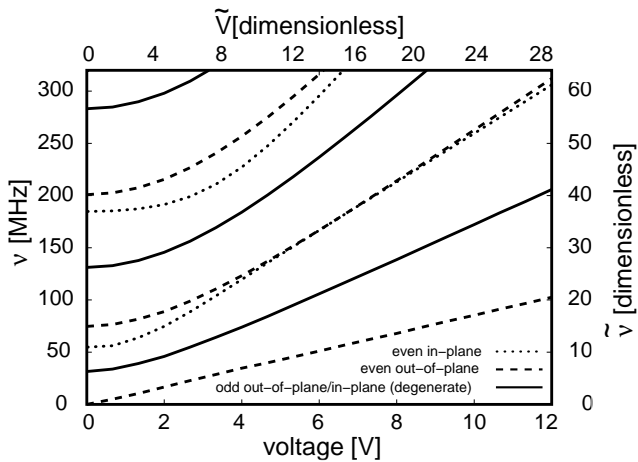}} \\
    (a) \\ 
    \scalebox{0.30}{\includegraphics{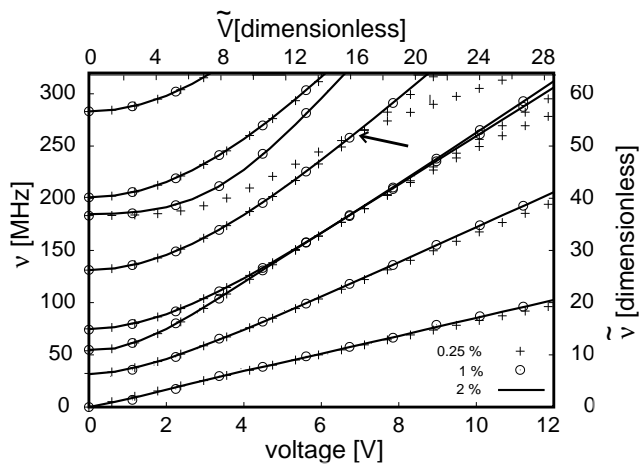}} \\
    (b) \\
  \end{tabular}
\end{center}
  \caption{(a) Dimensionless frequency versus force control parameter: in-plane
    modes (dashed curves), out-of-plane modes (dotted curves),
    degeneracy of in- and out-of- plane modes (solid curves). (b)
    Dimensionless frequency versus force control parameter scaled by
    fourth-root of slack ($v/\sqrt[4]{s/1\%}$) for three
    values of slack $s$: 0.25\% ('+'), 1\% (curves), 2\% ('$\odot$').
    Frequency in MHz and voltage in $V$ refer to the nanotube example
    described in the text. The scaled frequency $\tilde
    \nu=\nu\sqrt{\mu L^4/F}$ and the force control parameter $\tilde
    v=\sqrt{fL^3/F}$ apply to a general system.}
  \label{fig:intro}
\end{figure}

As a dimensionless measure of this force, we choose a {\em force control
parameter} $\tilde V\equiv \sqrt{fL^3/F}$, proportional to the square root
of the force.  In recent experiments on suspended nanotubes\cite{Nature},
the external force was introduced via a gate voltage, as shown in
Figure~\ref{fig:intro}a.  Such a voltage results in a force per unit length
$f=\frac{1}{2}\frac{dc}{dz}V^2$.  Here $c$ is the capacitance per unit
length of the nanotube given by
\begin{equation}
c=\frac{4\pi\epsilon_0}{2 \ln(z/d)},
\label{convert}
\end{equation}
where $\epsilon_0$ is the dielectric constant of vacuum and $z$ is the
distance between the wire and gate.  For the concrete example of a
$L$=1.75$\mu$m nanotube with a diameter of 2nm, we take the tube to be
suspended 500nm above the gate. In this case, a dimensionless parameter of
$\tilde V=10$ corresponts to a voltage of $\sim$4V.  As another example, a
100$\mu$m-long Si nanowire of 100nm diameter, with a $z$ of 25$\mu$m would
require a voltage of $\sim$60V to attain the same force control parameter
$\tilde V$ of ten.  Moving to larger systems, a 0.6mm-diameter steel piano
wire which is $\sim$2m long under the influence of gravity would have
the same dimensionless parameter of $\tilde V=10$.

\begin{figure}
\centering
\scalebox{0.30}{\includegraphics{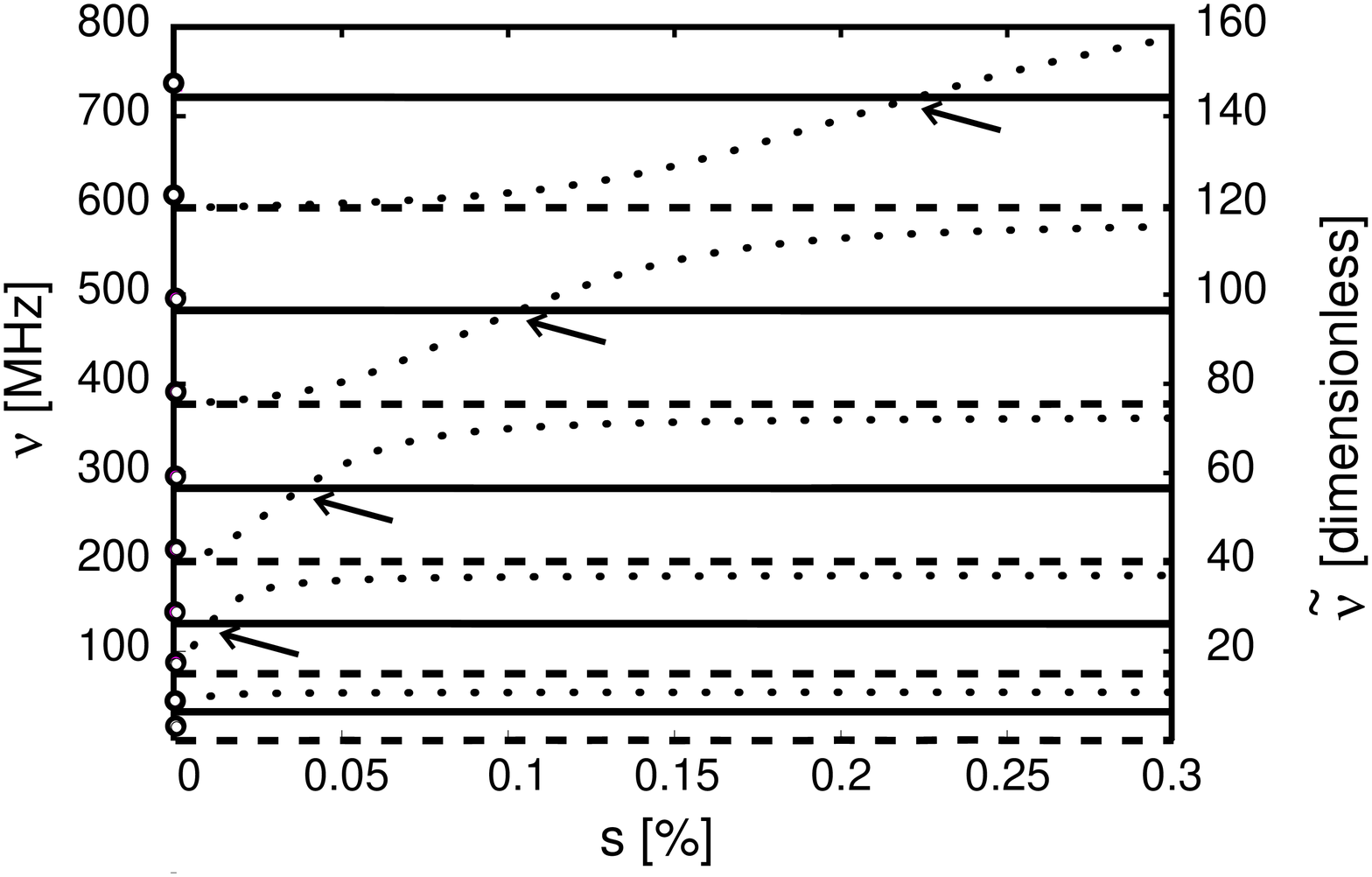}} \\
\caption{Frequency versus slack at zero force control parameter: classic
  solution for zero slack (points on vertical axis), numerical results for
  non-zero slack (curves).  The dimensionless frequency
  $\tilde\nu=\nu\sqrt{\mu L^4/F}$ applies to a general system.  The modes
  are doubly-degenerate and odd (solid curves), out-of-plane and even
  (dashed curves) or in-plane and even (dotted curves).  Arrows indicate
  mode crossings.
}
\label{fig:f_vs_s_0}
\end{figure}

Figures~\ref{fig:f_vs_s_0}~and~\ref{fig:intro} display our primary results,
both in dimensionless form and for the nanotube example, for the dependence
of the vibrational frequency on slack and force.  Figure~\ref{fig:f_vs_s_0}
shows the dependence of frequency on slack over the range of small slacks
for zero force.  The style of the curves indicate whether the mode is
doubly-degenerate and odd (solid), out-of-plane and even (dashed) or
in-plane and even (dotted).  After the mode crossings shown with arrows in
Figure~\ref{fig:f_vs_s_0}, all curves remain flat up through a slack of
2\%, the largest slack considered in this paper. Figure~\ref{fig:intro}
shows the dependence of the frequency on the force control
parameter. Figure~\ref{fig:intro}(a) shows the symmetry of the modes as in
Figure~\ref{fig:f_vs_s_0} for a system of 1\%
slack. Figure~\ref{fig:intro}(b) shows slacks of $s$=0.25\%, 1\% and 2\%
with the horizontal axis scaled by $\sqrt[4]{s}$ which, for reasons
described below, causes the curves to collapse upon each other for all but
the smallest slack.


\begin{figure}
  \begin{center}
    \scalebox{0.50}{\includegraphics{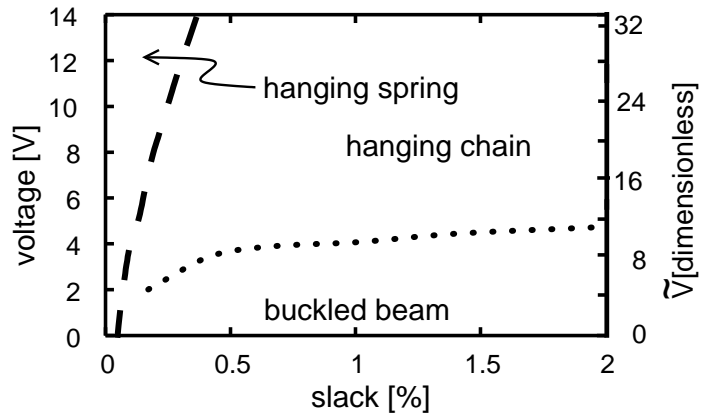}}
  \end{center}
  \caption{The three regimes in which nanotube behavior simplifies, as a
    function of voltage and slack.  The dashed line between the ``hanging
    spring'' and ``hanging chain'' regimes is the crossing between the
    second even in-plane mode and the second odd in-plane mode.  The dotted
    line separating the ``buckled beam'' regime from the ``hanging chain''
    regime is the contour on which the derivative $\frac{d\log\omega}{d\log
    s}$ has a value of $-1/8$. The axis on the right shows the force
    control parameter, $\tilde V=\sqrt{fL^3/F}$}
  \label{fig:regimes}
\end{figure}

We find three simplifying limits for the physical
behavior of the system as a function of slack $s$ and force control
parameter $v$.  In each such limit, a different term in the potential
energy (Equation~\ref{energy}) dominates with the other terms either being
negligible or acting as a constraint.  These three regimes, which we term
\emph{buckled beam}, \emph{hanging chain} and \emph{hanging spring}, are
shown in Figure~\ref{fig:regimes}.

For small external forces, the extensional rigidity acts as a constraint
which keeps the tube length fixed.  If the force is sufficiently small, the
bending term in Equation~\ref{energy} dominates the dynamics and the system
acts like an \emph{Euler-buckled beam}.  The \emph{in-plane} vibrations of
buckled beams have been studied analytically, but only in the absence of an
applied force and only under conditions of very small slack (under
$\sim0.07$\%~\cite{AIAA}).  Our results extend this treatment to three
dimensions and larger values of slack, probing at the same time the
dependence on the downward force.  In the absense of applied force
all modes are doubly degenerate at zero slack
with the frequencies of the classic unbuckled clamped beam~\cite{Landau},
shown as circles on the vertical axis of Figure~\ref{fig:f_vs_s_0}.

With the introduction of slack, the lowest out-of-plane mode, which we term
``jump rope'', corresponds to rotation of the relaxed system about the
clamping points, as illustrated in Figure~\ref{fig:experiment}b.
Rotational symmetry requires that this mode has zero frequency in the
absence of an external force.  It therefore shows up as a dashed line
overlaying the slack axis of Figure~\ref{fig:f_vs_s_0}. No such symmetry
applies to other out-of-plane modes, whose frequencies drop slightly, and
then remain essentially independent of slack.

In contrast to the out-of-plane modes, the in-plane modes are affected by a
length-conservation constraint which changes with the amount of slack.  For
reasonably small slack, the odd modes (shown as solid lines) are unaffected
by this constraint, since they conserve length by symmetry, and thus remain
degenerate with their out-of-plane counterparts.  The even modes (shown as
dotted lines), in order to conserve the tube length, acquire two extra
nodes as slack is introduced, leading to a mode crossing between the even
and odd modes (shown with arrows in Figure~\ref{fig:f_vs_s_0}).

When the force is large enough to overcome the bending term in
Equation~\ref{energy} but still too small to stretch the nanotube, the tube
behaves as a \emph{hanging chain} clamped at both ends, whose equilibrium
profile forms the well-known \emph{catenary}. In this regime, the slack
dependence of the frequency can be solved for analytically in the limit of
small slack.  In the small slack limit, the shape of the tube is parabolic,
and the tension is constant throughout the tube.  The frequency is then
determined by the result for a simple string under tension, giving
\begin{equation}
\omega_n=n\pi\sqrt{\frac{f}{\mu L\sqrt{24s}}}
\label{eq:catenary}
\end{equation}
Since in this equation the frequency is proportional to $s^{-1/4}$, we
scale the horizontal axis in Figure~\ref{fig:intro}(b) by $\sqrt[4]{s}$,
causing the curves for each slack value to collapse for small forces,
although for 0.25\% slack there is deviation in the higher modes, in
particular in the mode pointed to by the arrow.

For a given force, as the slack decreases, the tension increases and
eventually the tube begins to stretch and departs from the hanging chain
regime. This avoids the infinite frequency at very small slack which is
predicted by Equation~\ref{eq:catenary}. This effect is responsible for the
deviation from linear behavior of the tube with 0.25\% slack in
Figure~\ref{fig:intro}, in contrast to the tubes with 1\% and 2\% slack.

In the large force, small slack regime, the extensional term in
Equation~\ref{energy} can no longer be treated as a constraint, as the tube
begins to stretch.  The nanotube then behaves as a \emph{hanging spring}
clamped at both ends.  The frequency in this regime can be derived in a
manner similar to that in the hanging chain regime: the shape of the
nanotube is parabolic and the tension is constant throughout the
tube. However, in the case of the hanging spring, the tension is
proportional to the change in length of the tube.  Under these conditions,
the frequency is proportional to the cube root of the force:
\begin{equation}
\omega_n= n\pi \left(\frac53\right)^{1/3}
           \sqrt{\frac{{(EL)}^{1/3}}{6\mu L}}
           \quad f^{1/3}
\label{freq_str}
\end{equation}
The frequency is independent of slack, because as the tube is
stretched the original amount of slack in the nanotube becomes
insignificant.  However, the point of onset of the hanging spring
regime does depend on slack.

In the buckled beam and hanging chain regimes, the length conservation
constraint in a slack tube forces the first even in-plane mode to have two
nodes.  In contrast, in the hanging spring regime there is no length
conservation constraint, and consequently, the lowest frequency in-plane
mode has no nodes.  As the transition between these regimes occurs, there
is a mode crossing between the even and odd in-plane modes, shown with an
arrow in Figure~\ref{fig:intro} in the curve for 0.25\% slack.  The
crossing between the second even in-plane mode and the second odd in-plane
mode is shown as the dashed curve in Figure~\ref{fig:regimes} which
separates the hanging spring regime from the hanging chain regime.

In conclusion, we have studied numerically and analytically, as a function
of both the amount of slack and applied force, the transverse vibration
frequencies of one dimensional suspended systems clamped at both ends. We
have chosen as a concrete application to focus on suspended nanotubes under
the influence of a gate voltage which have recently been studied
experimentally~\cite{Nature}.  We find three regimes in which the behavior
simplifies and can be understood using approximate analytic solutions.
This understanding is important in characterizing and designing
nanoresonators where it can prove difficult to control or measure
parameters such as slack.  The hanging chain regime is particularly
suitable for tuning and characterizing nano-oscillators, since
Equation~\ref{eq:catenary}, which governs the frequency in this regime,
remains valid independently of imperfections affecting the boundary
conditions at the clamps.

This work was supported by the NSF through the Cornell Center for
Materials Research and the NIRT program.

\bibliographystyle{prsty}
\bibliography{paper}
\end{document}